\documentclass[preprint,12pt]{elsarticle} 
\usepackage{epsfig}
\usepackage{bm}

\begin{document}

\begin{frontmatter}

\title{Chaotic background of large-scale climate oscillations}

\author{ A. Bershadskii}

\address{ICAR - P.O. Box 31155, Jerusalem 91000, Israel}

\begin{abstract}
It is shown that the periodic alteration of night and day provides a chaotic dissipation 
mechanism for the North Atlantic (NAO) and Southern (SOI) climate oscillations.
The wavelet regression detrended daily NAO index for last 60 years and daily SOI for last 20 
years as well as an analytical continuation in the complex time domain were used for this purpose.

\end{abstract}

\end{frontmatter}

\begin{flushright}
"And in the alteration of night and day,~~

and the food We send down from the sky, 

and revives therewith the earth after its~~ 

death, and in the turning about of winds~ 

are signs for people of understanding".~~~~

QURAN (45:4-5).

\end{flushright}

\section{Introduction}

 It is now well known that climate is a 
nonlinear system. The nonlinear systems can exhibit a chaotic behavior. 
Both stochastic and deterministic processes can result in the
broad-band part of the spectrum, but the decay in the
spectral power is different for the two cases. The {\it exponential}
decay indicates that the broad-band spectrum
for these data arises from a deterministic rather than a
stochastic process (cf. Figs. 1-4). For a wide class of deterministic
systems a broad-band spectrum with exponential decay is a generic 
feature of their chaotic solutions (Ohtomo et al. 1995; Farmer 1982; 
Sigeti 1995; Frisch and Morf 1981). Also response of the chaotic systems 
to the periodic forcing does not always have the result that one might expect. 
Unlike linear systems, where periodic forcing 
leads to a periodic (peak-like) response, the chaotic response to periodic 
forcing can result in modification of the rate of the exponential decay of the
broad-band power spectrum. Thus, the chaotic response to periodic forcing can utilize
the generic dissipation mechanism for the chaotic dissipative systems, especially in the case when 
the system's fundamental (internal) frequency is considerably lower than the frequency 
of the periodic forcing under consideration.  

  The climate, where the chaotic behavior was discovered for the first time, is still one of the 
most challenging areas for the theory of chaos. 
The weather (time scales up to several weeks) chaotic behavior usually can be directly 
related to chaotic convection, while appearance of the 
chaotic properties for more long-term climate events is a non-trivial and 
challenging phenomenon.

   The solar day is a period of time during which the earth makes one revolution 
on its axis relative to the sun. During a part of the day the sun's direct rays are 
blocked (locally) by the earth. Therefore, the periodic {\it daily} variability of solar impact plays 
crucial role in high frequency climate behavior. It will be shown in present paper that just unusual properties of chaotic response to the daily periodicity of solar forcing provide an effective mechanism
for high frequency dissipation of the large-scale climate oscillations, presumably generated by the instabilities related to large scale topography. 

\section{Large-scale climate oscillations} 
 
 One of the most significant and recurrent patterns of atmospheric variability over the middle and high 
latitudes of the Northern Hemisphere is known as NAO - North Atlantic Oscillation. 
Climate variability from the subtropical Atlantic to the Arctic and from Siberia to the eastern 
boards of the North America is strongly related to the NAO (see, for a comprehensive review Hurrell et al. 
(2003)). In site NAO (2010a) a projection of the daily 500mb 
height anomalies over the Northern Hemisphere onto the loading pattern (see also NAO, 2010b) of the NAO 
was used in order to construct the daily NAO for last 60 years. Due to the natural climatic trends the daily NAO 
index time series is not a statistically stationary data set. In order to solve this problem 
a wavelet regression detrending method (Ogden, 1997) was used in present investigation for the daily NAO 
time series. We used a symmlet regression 
of the data. Most of the regression methods are linear in responses. 
At the nonlinear nonparametric wavelet regression one chooses a relatively small number of wavelet 
coefficients to represent the underlying regression function. A threshold method is used to keep or 
kill the wavelet coefficients. In this case, in particular, the Universal (VisuShrink) thresholding 
rule with a soft thresholding function was used. At the wavelet 
regression the demands to smoothness of the function being estimated are relaxed considerably in comparison 
to the traditional methods. 
Figure 1 shows a spectrum of the wavelet regression detrended data (NAO, 2010a) calculated using the maximum 
entropy method. This method provides an optimal spectral resolution even for small data sets. 
The spectrum exhibits a broad-band behavior with exponential decay: 
$$
E(f) \sim e^{-4\pi f}   \eqno{(1)}
$$
A semi-logarithmic plot was used in Fig. 1 in order to
show the exponential decay (at this plot the
exponential decay corresponds to a straight line, for an explanation 
see below). 

\begin{figure} \vspace{-3cm}\centering
\epsfig{width=.8\textwidth,file=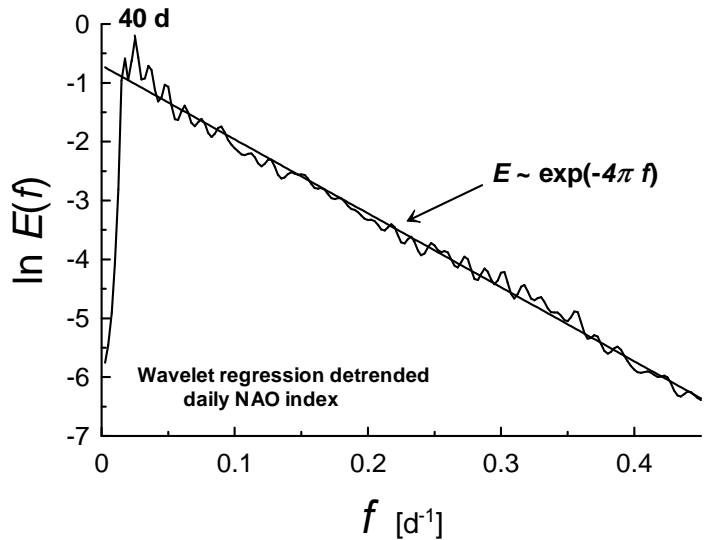} \vspace{-6.5cm}
\caption{Spectrum of the wavelet regression detrended NAO data series 
(the NAO index data were taken from the site NAO, 2010a). In the semi-logarithmic
scales used in the figure the straight line indicates the exponential
decay Eq. (1).}
\end{figure}

Another example of the large-scale climate oscillations with the chaotic dissipation mechanism 
can be recognized in Tropical Atmosphere, where one of the most significant and recurrent patterns of
atmospheric variability between the western and eastern tropical Pacific is known 
as Southern Oscillation. The Southern Oscillation is related to the variability of the 
Walker circulation system: a circulation pattern characterized 
by sinking air above the eastern Pacific and rising air above the western Pacific. 
The Southern Oscillation is often considered as the atmospheric component of 
$El~Ni\tilde{n}o$ phenomenon. The daily Southern Oscillation Index (SOI) 
has been calculated based on the differences in air pressure anomaly between Tahiti and Darwin, 
Australia (Troup, 1965) for last 20 years (SOI, 2010). 
Figure 2 shows a spectrum of the wavelet regression detrended data calculated using the 
maximum entropy method. A semi-logarithmic plot was used in 
Fig. 2 in order to show the exponential decay Eq. (1) (cf Fig.1). 

\begin{figure} \vspace{-3cm}\centering
\epsfig{width=.8\textwidth,file=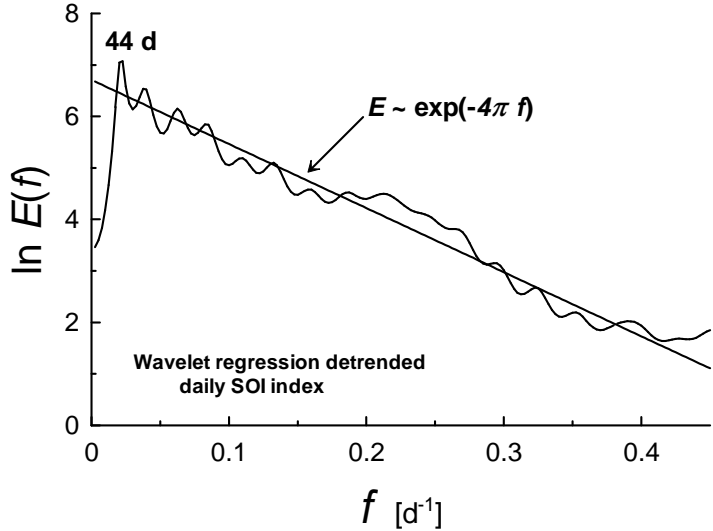} \vspace{-6.5cm}
\caption{Spectrum of the wavelet regression detrended SOI data series 
(the SOI data were taken from the site SOI, 2010). In the semi-logarithmic
scales used in the figure the straight line indicates the exponential
decay Eq. (1).}
\end{figure}

\section{Chaos and exponential spectrum}

For a wide class of deterministic
systems a broad-band spectrum with exponential
decay is a generic feature of their chaotic solutions (Ohtomo et al. 1995; 
Farmer 1982; Sigeti 1995; Frisch and Morf 1981). Let us consider a relevant 
example. The Lorenz equations are given by: 
$$
\frac{dx}{dt} = \sigma (y - x),~~      
\frac{dy}{dt} = r x - y - x z, ~~
\frac{dz}{dt} = x y - b z      \eqno{(2)}          
$$
The standard values producing a chaotic attractor are: 
$\sigma=10.0,~ r = 28.0,~ b = 8/3$. Figure 3 shows a power 
spectrum for $z$-component of the Lorenz chaotic attractor 
generated by Eq. (2) (the spectrum was again 
calculated using the maximum entropy method). 
A semi-logarithmic plot was used again in Fig. 2 in order to
show exponential decay (at this plot the
exponential decay corresponds to a straight line).

\begin{figure} \vspace{-3cm}\centering
\epsfig{width=.8\textwidth,file=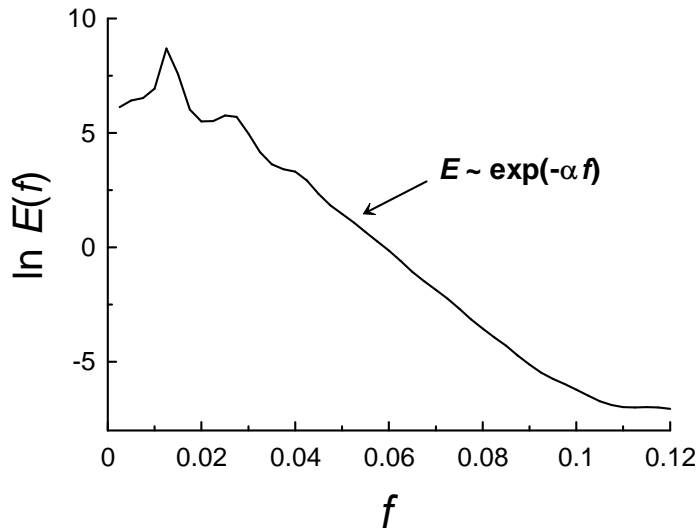} \vspace{-6cm}
\caption{As in Fig.1 but for $z$-component of the Lorenz chaotic attractor Eq. (2). 
In the semi-logarithmic scales used in the figure the straight line indicates 
the exponential decay.}
\end{figure}

 Another relevant example is a modulated Duffing oscillator. The driven Duffing oscillator 
has become a classic model for
analysis of nonlinear phenomena and can exhibit both deterministic and chaotic behavior: Ott, 2002; 
Permann and Hamilton, 1992). It is well known that for strong nonlinearity 
a driven Duffing oscillator can exhibit chaotic properties and, in particular, 
exponentially decaying broad-band spectrum (see, for instance, Ohtomo et al. (1995)). 
In the case of a weak nonlinearity, however, one needs in an 
additional slow modulation of the driving force in order to obtain a chaotic behavior. 
In this case a separation between fast and slow motion is possible and in certain range of 
parameters the slow component exhibits chaotic properties with an exponentially decaying broad-band 
spectrum (see, for instance, Miles (1984)). The driven Duffing oscillator is described by the equation
$$
\ddot{x}+ \omega_0^2 x(1+\eta x^2)+ 2\delta \omega_0 \dot{x} = A(t) cos(\omega_c t)    \eqno{(3)}
$$
where $\dot{x}$ denotes the temporal derivative of $x$, $\delta$ is a damping parameter, 
$\eta$ is a nonlinearity parameter, $\omega_0$ is the natural frequency for free oscillations, 
$\omega_c$ is the carrier frequency. The amplitude $A(t)$ can be slow modulated. Namely, 
$A(t) = \varepsilon \omega_0^2 cos(\omega_m t + \theta_0)$, where $\omega_m$ is a modulation frequency, 
$\theta_0$ is a phase constant, $\varepsilon$ is a small constant amplitude. 
Except of a resonant domain: $(\omega_c^2-\omega_0^2)/\omega_0^2$ and $\omega_m/\omega_0= O(\varepsilon^{2/3})$ 
the nonlinearity is unimportant in the limit $\varepsilon \longrightarrow 0$. 
Following to Miles (1984) one can introduce a slow time $\tau = \frac{1}{2}\sigma \omega_c t$
and pose the solution of Eq. (3) in the form (Van der Pol transformation):
$$
x(t)= \frac{\varepsilon}{\sigma} [u_1(\tau) cos(\omega_c t) + u_2(\tau) sin(\omega_c t)]  \eqno{(4)}
$$
where $u_1(\tau)$ and $u_2(\tau)$ are slowly varying amplitudes, $\sigma \sim O(\varepsilon^{2/3})$ is 
an arbitrary dimensionless positive constant. 

\begin{figure} \vspace{-5cm}\centering
\epsfig{width=.8\textwidth,file=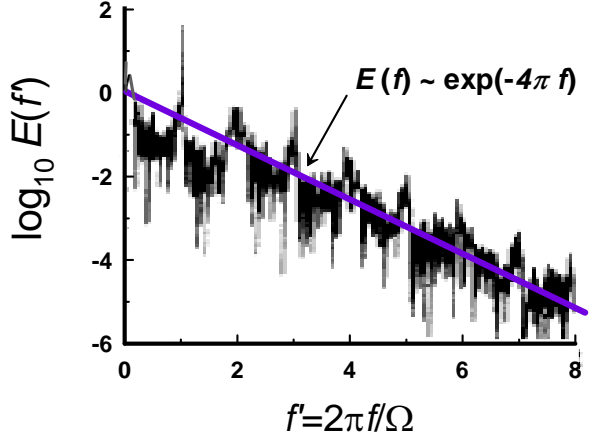} \vspace{-7cm}
\caption{As in Fig.3 but for slow component of the modulated Duffing oscillator. 
In the semi-logarithmic scales used in the figure the straight line indicates 
the exponential decay. The data were taken from Miles (1984).}
\end{figure}

Then the equations for the slow components are  
$$
\frac{du_1}{d\tau}=U_1(u_1,u_2),~~~~~~~~\frac{du_2}{d\tau}= U_2 (u_1,u_2) + cos \theta   \eqno{(5)}
$$
where $\theta= \Omega \tau + \theta_0$, $\Omega= 2\omega_m/\sigma \omega_c$, and 
$U_1(u_1,u_2)$ and $U_2(u_1,u_2)$ 
are certain linear functions on $u_1$ and $u_2$ (see for more details Miles (1984)). 
Numerical analysis of the equations 
(5) performed by Miles (1984) showed that for certain values of the parameters the slow component of the 
oscillations exhibits chaotic properties. In particular, the broad-band spectra with exponential decay were observed  
by Miles (1984) for these values of parameters. Figure 4 shows an example of such spectrum ($\Omega =0.75$, 
the data were taken from Miles (1984)). The spectra were calculated by Miles (1984) using 
a fast-Fourier-transform with frequency $f \longleftrightarrow \tau$. One can see a clear peak corresponding to the fundamental frequency $f_{fun} = \Omega/2\pi$. We have drawn the straight line in Fig. 4 in order to indicate the exponential decay: Eqs. (1). The same exponential decay - Eq. (1), can be observed for all 'chaotic' values of the 
parameter $\Omega$, i.e. it has an universal nature.  
 
  Nature of the exponential decay of the power spectra
of the chaotic systems is still an unsolved mathematical
problem. A progress in solution of this problem
has been achieved by the use of the analytical continuation
of the equations in the complex domain (see, for 
instance, Frisch and Morf (1981)). In this approach the exponential decay
of chaotic spectrum is related to a singularity in the
plane of complex time, which lies nearest to the real axis.
Distance between this singularity and the real axis determines
the rate of the exponential decay. For many interesting cases 
chaotic solutions are analytic in a finite strip around the real time axis. 
This takes place, for instance for attractors bounded in the real 
domain (the Lorenz attractor, for instance). 
In this case the radius of convergence of the Taylor series 
is also bounded (uniformly) at any real time. Let us consider, for 
simplicity, solution $u(\tau)$ with simple poles only, and to define the Fourier 
transform as follows
$$
u(\omega) =(2\pi)^{-1/2} \int_{-T_e/2}^{T_e/2} d\tau~e^{-i \omega t} u(\tau)  \eqno{(6)}
$$  
where the function $u(\tau)$ is defined on the interval $-T_e/2 < \tau < T_e/2$ with 
periodic boundary conditions. Then using the theorem of residues
$$
u(\omega) =i (2\pi)^{1/2} \sum_j R_j \exp (i \omega x_j -|\omega y_j|)  \eqno{(7)}
$$
where $R_j$ are the poles residue and $x_j + iy_j$ are their location in the relevant half
plane of the complex time, one obtains asymptotic behavior of the spectrum 
$E(\omega)= |u(\omega)|^2$ at large $\omega$
$$
E(f) \sim \exp (-4\pi f~ y_{min})  \eqno{(8)}
$$
where $\omega= 2\pi f$ and $y_{min}$ is the imaginary part of the location of
the pole which lies nearest to the real axis. 

 If in the considered case $y_{min}=1$d, then 
we obtain the exponential decay shown in Fig. 1 (cf Eq. (1)). In order to understand the reason for 
$y_{min}=1$d we need in a nonlinear dynamic climate model where the {\it daily} periodic solar impact results 
in such position of the pole nearest to the real axis. But even before constructing such dynamic model 
it is already clear that just the {\it daily} periodicity of the solar impact 
is responsible for the chaotic exponential decay shown in Fig. 1. 
\begin{figure} \vspace{-3cm}\centering
\epsfig{width=.8\textwidth,file=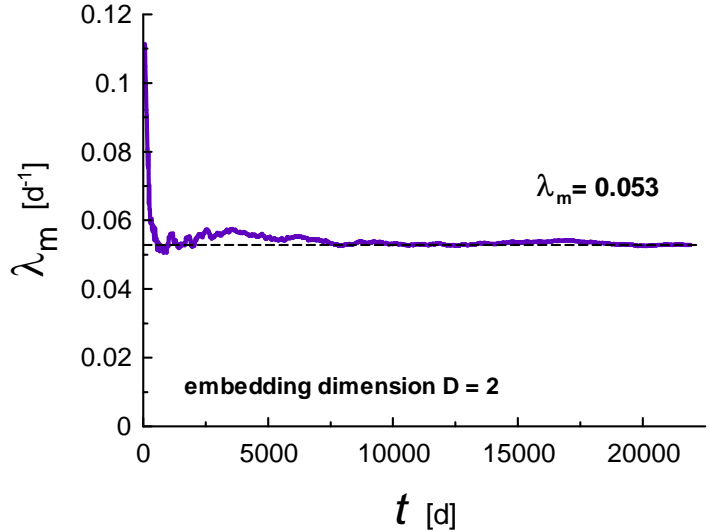} \vspace{-6cm}
\caption{The pertaining average maximal Lyapunov exponent 
at the pertaining time, calculated for the same data 
as those used for calculation of the spectrum (Fig. 1). 
The dashed straight line indicates convergence to a positive value.}
\end{figure}

  A confusion can arise due to the 1 day period also being the sampling frequency for the NAO and SOI time series. However, the poles location in the complex time plane is presumably determined by the system's {\it dynamics} and not by the sampling rate. For the considered above examples, for instance, the exponential decay rate (Figs. 3 and 4) 
is not equal to $4\pi s$, $~s$ being the sampling rate for the corresponding time series, as it would be otherwise.   \\

Additionally to the exponential spectrum, let us check the chaotic character of the wavelet regression detrended 
NAO data set calculating the largest Lyapunov exponent : $\lambda_{max}$. A strong indicator for the presence 
of chaos in the examined time series is condition $\lambda_{max} >0$. If this is the case, then 
we have so-called exponential instability. Namely, 
two arbitrary close trajectories of the system will diverge apart exponentially, that is 
the hallmark of chaos. To calculate $\lambda_{max}$ we used a direct algorithm developed by 
Wolf et al. (1985), Kodba et al. (2003). Figure 4 shows 
the pertaining average maximal Lyapunov exponent at the pertaining time, calculated for the same data 
as those used for calculation of the spectrum (Fig. 1). The largest Lyapunov exponent converges very 
well to a positive value $\lambda_{max} \simeq 0.053 d^{-1} > 0$. \\
 
\section{Conclusions}

 The exponential decay of the broad-band spectrum of the wavelet regression detrended NAO and SOI 
time series (Figs. 1 and 2) indicates a chaotic nature of the high-frequency components of these climate 
oscillations.  In Fig. 1 one can readily recognize a peak corresponding to a dominant frequency of 
internal variability (corresponding to a period approximately equal to 40 days, cf. with the fundamental 
frequency of the chaotic systems: Figs. 3 and 4). It should be noted, 
that the 40-day oscillations are well known as an intrinsic mode of the Northern Hemisphere extratropics, 
which is exited presumably due to instabilities related to large-scale topography 
(see, for instance, Magana 1993; Marcus, Ghil and Dickey 1994; Ghil and Robertson 2002). 
In this case the dominant frequency of internal variability (corresponding to the period $\sim$40d) 
is considerably smaller than the daily solar forcing frequency, which determines the rate of the spectral 
exponential decay (Eq. (8)). Thus, the exponential chaotic decay (Fig. 1) can be considered as a 
dissipation mechanism for these oscillations. Figure 2 shows analogous picture for the SOI data.

\section{Acknowledgments}

I thank to J. Hurrell and to the Department of Environment and Resource Management 
(Queensland) for sharing their data.

\newpage 

\begin{center}

{\bf References}

\end{center}

Farmer J.D, 1982, Chaotic attractors of an infinite dimensional dynamic system, Physica D, {\bf 4}, 
366-393.

Frisch U. and R. Morf, 1981, Intermittency in non-linear dynamics
and singularities at complex times, Phys. Rev., {\bf 23}, 2673-2705.

Ghil M., and A.W. Robertson, 2002, "Waves" vs. "particles" in the atmosphere's phase space: 
A pathway to long-range forecasting? Proc. Natl. Acad. Sci., {\bf 99} (Suppl. 1), 2493-2500.

Hurrell J.W. et al., 2003, An Overview of the North Atlantic Oscillation, in 
"The North Atlantic Oscillation: Climatic Significance and Environmental Impact", 
Geophysical Monograph {\bf 134}, p.1, American Geophysical Union (2003).

Kodba S., Perc M., and Marhl M., 2005, Detecting chaos from a time series,  
Eur. J. Phys. {\bf 26}, 205-215.

Magana V., 1993, The 40-day and 50-day oscillations in atmospheric angular 
momentum at various latitudes, J. Geophys. Res., {\bf 98}, 10441-10450.

Marcus S.L., Ghil M., and J.O. Dickey, 1994, The extratropical 40-day oscillation in the 
UCLA General Circulation Model, Part I: Atmospheric angular momentum, J. Atmos. Sci., 
{\bf 51}, 1431-1466.

Miles J. 1984, Chaotic motion of a weakly nonlinear, modulated oscillator, PNAS, {\bf 81}, 
3919-3923.

NAO, 2010a, http://www.cgd.ucar.edu/cas/jhurrell/indices.html

NAO, 2010b, http://www.cpc.ncep.noaa.gov/products/precip/CWlink

/pna/nao\_loading.html

Ogden T., 1997, Essential Wavelets for Statistical Applications and Data Analysis, 
Birkhauser, Basel.

Ohtomo N. et al., 1995, Exponential Characteristics of Power Spectral Densities Caused by Chaotic Phenomena, 
J. Phys. Soc. Jpn., {\bf 64}, 1104-1113.

Ott  E., 2002. Chaos in Dynamical Systems (Cambridge University Press).

Permann D. and I. Hamilton, 1992. Wavelet analysis of time series for the Duffing oscillator: 
The detection of order within chaos, Phys. Rev. Lett., {\bf 69}, 2607-2610.

Sigeti D.E., 1995,  Survival of deterministic dynamics in the presence of noise and the
exponential decay of power spectrum at high frequencies. Phys. Rev. E, {\bf 52}, 2443-2457.

SOI, 2010, http://www.longpaddock.qld.gov.au/SeasonalClimate 

Outlook/SouthernOscillationIndex/SOIDataFiles/index.html

Troup A.J., 1965, The Southern Oscillation. Quart. J. Roy. Met. Soc., {\bf 91}, 490-506.

Wolf A. et al., 1985, Determining Lyapunov exponents from a time series, 
Physica D, {\bf 16}, 285-317.

\end{document}